\documentclass[conference]{IEEEtran}
\usepackage{epsfig}
\usepackage{times}
\usepackage{float}
\usepackage{afterpage}
\usepackage{amsmath}
\usepackage{amstext}
\usepackage{amssymb,bm}
\usepackage{latexsym}
\usepackage{color}
\usepackage{array,algpseudocode}
\usepackage{makecell}
\usepackage{diagbox}
\usepackage{amsmath}
\usepackage{amsthm}
\usepackage{graphicx}
\usepackage[center]{caption}
\usepackage{pstricks}
\usepackage{subfigure}
\usepackage{booktabs}
\usepackage{enumerate}
\usepackage[ruled]{algorithm}
\usepackage{romannum}
\usepackage{derivative}
\usepackage{cite}
\usepackage{lipsum}
\usepackage{cuted}
\usepackage{stackengine}
\def\delequal{\mathrel{\ensurestackMath{\stackon[1pt]{=}{\scriptstyle\Delta}}}}

\usepackage{bbm}
\def\BibTeX{{\rm B\kern-.05em{\sc i\kern-.025em b}\kern-.08em
    T\kern-.1667em\lower.7ex\hbox{E}\kern-.125emX}}
\usepackage[normalem]{ulem}
\usepackage{url}
\newtheorem{theorem}{Theorem}
\newtheorem{lemma}{Lemma}
\newtheorem{definition}{Definition}
\newtheorem{remark}{Remark}
\newtheorem{example}{Example}

\newtheorem{proposition}{Proposition}

\newcommand{\comment}[1]{}
\newcommand{\floor}[1]{\lfloor #1 \rfloor}
\newcommand{\ceil}[1]{\lceil #1 \rceil}
\let\svthefootnote\thefootnote
\newcommand\blankfootnote[1]{%
  \let\thefootnote\relax\footnotetext{#1}%
  \let\thefootnote\svthefootnote%
}

\allowdisplaybreaks

\begin{document}
\title{On the State Estimation Error of ``Beam-Pointing'' Channels: The Binary Case}
\author{\IEEEauthorblockN{Siyao Li}
\IEEEauthorblockA{\textit{ Communications and Information Theory Group (CommIT) } \\
\textit{ Technische Universit\"{a}t  Berlin }\\
Berlin, Germany \\
siyao.li@tu-berlin.de}
\and
\IEEEauthorblockN{Giuseppe Caire}
\IEEEauthorblockA{\textit{Communications and Information Theory Group (CommIT)} \\
\textit{ Technische Universit\"{a}t Berlin }\\
Berlin, Germany  \\
caire@tu-berlin.de} }
\maketitle

\begin{abstract}
Sensing capabilities as an integral part of the network have been identified as a novel feature of sixth-generation (6G) wireless networks. As a key driver, millimeter-wave (mmWave) communication largely boosts speed, capacities, and connectivity. In order to maximize the potential of mmWave communication, precise and fast beam acquisition (BA) is crucial, since it compensates for a high pathloss and provides a large beamforming gain. Practically, the angle-of-departure (AoD) remains almost constant over numerous consecutive time slots, the backscatter signal experiences some delay, and the hardware is restricted under the peak power constraint.  
This work captures these main features by a simple binary beam-pointing (BBP) channel model with in-block memory (iBM)~\cite{Kramer-memory}, peak cost constraint, and one unit-delayed feedback. In particular, we focus on the sensing capabilities of such a model and characterize the performance of the BA process in terms of the Hamming distortion of the estimated channel state. We encode the position of the AoD and derive the minimum distortion of the BBP channel under the peak cost constraint with no communication constraint. Our previous work \cite{Li2022OnTC} proposed a joint communication and sensing (JCAS) algorithm, which achieves the capacity of the same channel model. Herein, we show that by employing this JCAS transmission strategy, optimal data communication and channel estimation can be accomplished simultaneously. This yields the complete characterization of the capacity-distortion tradeoff for this model.
\end{abstract}

\section{Introduction}
\label{sec:intro}
With the evolution of 4G to 5G, the spectrum allocations have expanded towards millimeter-wave (mmWave) bands \cite{OverviewmmWave}. This trend will continue and communication spectra in the sub-Terahertz region will likely be available as some of the frequency bands for 6G deployments. With the introduction of these new frequencies, the potential for very accurate sensing based on radar-like technology arises~\cite{5G-fixed-wireless-access, SensorIngelligence,Location-aware}.
  That is, reflections of transmitted signals are received in the network and processed to yield spatial knowledge of the physical surroundings. At these frequencies, the communication network must employ beamforming of the transmitted signals to concentrate and direct the signal energy to a specific geographical area where the intended receiver is located \cite{mmWave5G}, which is strongly affected by the {\it initial beam acquisition} (BA) phase \cite{BeamDetection}. In general, standard BA schemes are based on some ``beam-sweeping'' phase, i.e., the base station (BS) sends pilot signals in all the possible transmission directions at regular intervals, allowing the  user equipment (UE) to identify the best beam index and feed this back via some hand-shaking protocol. 
Works have studied the BA problem in various ways (e.g., see  \cite{Desai_2014,Javidi-AL,Hashemi_2018,Michelusi_2018,Scalable-song,mmPosterior,Hussain_2019} and references therein). However, transmission efficiency is not fully realized in these works, as they isolate the BA phase and the data communication phase. This separation is known to be sub-optimal from the information-theoretic perspective \cite{Lapidoth1998}. 

For future sensing, the main advantage of the communication network is that most of the infrastructure is already in place with transmitter/receiver (Tx/Rx) nodes. This provides full area coverage as well as a good interconnection between nodes. Hence, the sensing can be provided almost ‘for free’. To achieve the full potential of mmWave communication, numerous recent works exploit ``joint communication and sensing' (JCAS) (see e.g.  \cite{ISAC-6G,ISAC-Tradeoff,ISAC-MAC,ISAC-IT} and references therein), where communication can take place while the angle-of-departure (AoD) is being estimated via the backscatter signal.

 From physical considerations, it is clear that the AoD remains almost constant over a large number of consecutive time slots, which presents a state-dependent channel with memory. Additionally, 
 the backscatter signal can be modeled as causal feedback. 
 In this work, we investigate this scenario with JCAS from an information-theoretic viewpoint. 
Considering channels with {\em in-block memory} (iBM)~\cite{Kramer-memory}, i.e., the state remains constant for blocks of $L$ time slots, and changes in an independent and identically distributed (i.i.d.) fashion from block to block, the state estimation error is very hard to evaluate since it requires the optimization over length-$L$ sequences of conditional input distributions.

\paragraph*{Contribution}  
In this paper, we consider a binary beam-pointing (BBP) channel with iBM and one-unit delayed feedback under the peak cost constraint. 
In our previous work \cite{Li2022OnTC}, we characterized the capacity of this BBP channel. Herein, we are interested in the ability of the BS to “locate” the target AoD, quantified by the distortion error achieved by the BS state estimator at the end of each block. We refer to ``distortion" as the (average) error with which the transmitter is able to determine the channel state (i.e.,  the AoD of the receiver) at the end of each block, and characterize the minimum distortion under the peak cost constraint. It is interesting to see that this minimum distortion can be obtained by deploying the capacity-achieving transmission strategy in \cite{Li2022OnTC}. We, therefore, obtain a complete capacity-distortion region of the considered channel model, revealing that for this model the tradeoff is ``trivial’’ in the sense that optimal communication rate and minimum state estimation distortion can be achieved at the same time. This corroborates the general intuition that JCAS yields excellent sensing capabilities without compromising capacity. 
\paragraph*{Notations} 
For an integer $n$, we let $[n] = \{ 1, \cdots, n\}$ and $[n_1: n_2] = \{ n_1, \cdots, n_2\}$ for some integers $n_1 < n_2$.  $\underline{X}$ denotes a vector and $\underline{X}^n = [\underline{X}_1, \cdots, \underline{X}_n ]$ denotes a sequence of vectors. Let  $[x]^+ = \max(0, x)$, and $y^{i-1}1$ denote the realization of the sequence $Y^i$ where $Y^{i-1} = y^{i-1}$ and $Y_i=1$. Let $\beta_{k}^i$ denote the  binary sequence with  length $i$ and the first one element appearing at the $k$-index (i.e., $\beta_2^i = \{01 \underbrace{\star \cdots \star}_{i-2} \}$ where $\star$ can be either 0 or 1). 
$\mathbbm{1}{\{ \cdot \}}$ denotes an indicator function.  
 $| \mathcal A | $ represents the cardinality of a set $\mathcal A$. 

\section{System Model}
\label{sec:generalized}
We consider a BBP channel model with iBM~\cite{Kramer-memory}  
and block length $L$.  There are $n$ total transmission time in channel uses where $n =  \ell L$ and $\ell$ is the number of blocks. Note that when $L=1$, the channel becomes the memoryless channel with independent states.
The channel state ${\underline S} \in \{0, 1\}^M$ is an $M$-dimensional ``one-hot" binary vector where the single ``1" appearing at index $m$ indicates the (unknown) target receiver AoD, among $M$ possible quantized angles. This index $m$ is a  random variable uniformly distributed over $[M]$ and is referred to as the {\it transmission direction},  i.e., the quantized AoD of the UE with respect to the BS array. The state remains constant for blocks of $L$ channel uses and the transmitter (BS) receives binary causal noiseless feedback. This channel state information is assumed to be perfectly known at the receiver (CSIR) but unknown at the transmitter. The transmitter decides the transmission direction estimation $\hat{S}$ (i.e., a one-hot vector) upon the channel input and feedback at the end of each block. Furthermore, ${\underline S} \in \mathcal S$ is i) independent of the channel input, ii) remains constant for an interval of $L$ channel uses, and iii) i.i.d. according to $P_{\underline S}$ across the blocks. The channel input ${ \underline X}_{i,j} \in \mathcal X  := \{0,1\}^M$ is also an $M$-dimensional binary vector with a peak Hamming weight cost constraint, modeling the fact that sending in multiple directions costs transmit power. The channel output $Y_{i,j} \in \mathcal Y := \{0,1\} $  
at channel use $j$ of  block $i$
\begin{align}
Y_{i,j} = {\underline S}_i^T {\underline X}_{i,j} \label{eq:channel-model}
\end{align} 
is binary, given by the inner product of the state and input vectors. The causal feedback is noiseless, i.e., it coincides with the output from the previous channel use. 
Notice that $Y_{i,j} = 1$ if the single ``1’’ in $S_i$ coincides with a ``1’’ in $X_{i,j}$ and zero otherwise. 
The joint probability distribution of the considered model is 
\begin{align}
&P_{W {\underline X}^n {\underline S}^{\ell} Y^n }(  w, {\underline x}^n, {\underline s}^{\ell}, y^n) 
=P_W(w) \notag
\\&\times   \prod_{i=1}^{\ell} \left( P_{\underline S}( {\underline s}_i ) \prod_{j=1}^L P_{Y|{\underline X \underline S} }(  y_{i,j}  | {\underline x}_{i,j}  {\underline s}_i ) P({\underline x}_{i,j} | w, { y}_{i}^{j-1} ) \right) 
\label{eq:joint-distribution}
\end{align} 
where 
we denote $ y_{i}^{j-1}  = [y_{i,1}, \cdots, y_{i, j-1}]$.

\begin{definition}\normalfont
 The estimate of the state sequence $\mathcal S^{\ell}$ in the presence of the input $ X^n$ and feedback $Y^n$ is defined as
\begin{align}
 \hat{ \underline S}^{\ell} \delequal q(  \underline X^n, Y^n), \label{eq:hat-S}
\end{align} 
where 
$q: \mathcal X^n \times \mathcal Y^n \to \hat{\mathcal S}^{\ell},$
is a state estimation function and $\hat{\mathcal S}$ is the reproduction alphabet. 
The average per-block distortion is defined as
\begin{align}
\Delta^{(\ell)} \delequal \frac{1}{\ell} \sum_{i=1}^{\ell} \mathbb{E}[ d( \underline S_i, \hat{ \underline S}_i ) ], \label{eq:delta-ell}
\end{align}
where $\hat{\underline S}_i$ is the $i$-th component of  $\hat{ \underline S}^{\ell}$ in \eqref{eq:hat-S}, and 
$d: \mathcal S \times \hat{\mathcal S} \to \mathbb R_+$
is a state estimation error measure with $\max_{ ({ \underline s}, \hat{\underline s}) \in \mathcal S \times \hat{\mathcal S}} d({ \underline s}, \hat{\underline s} ) < \infty.$
\end{definition}
\begin{lemma} \label{lemma-s-star}
Define the function 
\begin{align*}
\hat{\underline s}^{\star}(\underline x_i^L, y_i^L) \delequal \arg \min_{\underline s_i^{\prime} \in \hat{\mathcal S}} \sum_{ \underline s_i \in \mathcal S} P_{ \underline S_i | \underline X_i^L Y_i^L}(\underline s_i| \underline x_i^L, y_i^L) d( \underline s_i,  \underline s_i^{\prime})
\end{align*}
where 
\begin{align*}
P_{ \underline S_i | \underline X_i^L  Y_i^L}(\underline s_i| \underline x_i^L,  y_i^L) = \frac{P_{ \underline S_i}(\underline s_i) P_{Y_i^L| \underline S_i, \underline X_i^L }(y_i^L| \underline s_i, \underline x_i^L)}{\sum_{ \underline s_i \in \mathcal S} P_{\underline S_i}( \underline s_i) P_{Y_i^L| \underline S_i, \underline X_i^L }(y_i^L|\underline s_i,  \underline x_i^L)},
\end{align*}
and 
$P_{Y_i^L|  \underline S_i, \underline X_i^L }(y_i^L| \underline s_i, \underline x_i^L) = \prod_{j=1}^L P_{Y_i |  \underline S_i,  \underline X_i }(y_{i,j} | \underline s_i, \underline x_{i,j})$.
Irrespective of the choice of encoding and decoding functions, distortion $\Delta^{(\ell)}$ in \eqref{eq:delta-ell} is minimized by the estimator
\begin{align*}
q^{\star}(\underline x^n, y^n) = ( \hat{\underline s}^{\star}(\underline x_1^L, y_1^L) ,  \hat{\underline s}^{\star}(\underline x_2^L, y_2^L)\cdots, \hat{\underline s}^{\star}(\underline x_{\ell}^L, y_{\ell}^L))
\end{align*} where $\hat{\underline s}^{\star}(\underline x_i^L, y_i^L)$ is the state estimation of the $i$-th block, $i\in[\ell]$.
\end{lemma}
\begin{IEEEproof}
By \eqref{eq:hat-S}, we have
\begin{subequations}
\begin{align}
&\mathbb{E}[d(  \underline S_i, \hat{ \underline S}_i)] \notag
 \\& = \mathbb{E}_{ \underline X^n, Y^n} \left[ \mathbb{E}[d( \underline S_i, \hat{\underline S}_i) |  \underline X^n, Y^n] \right] \notag
\\ & = \sum_{ \underline x^n, y^n} P_{ \underline X^n, Y^n}( \underline x^n, y^n) \sum_{\hat{\underline s}_i \in \mathcal S} P_{ \hat{\underline S}_i |  \underline X^n, Y^n} (\hat{ \underline s} |  \underline x^n, y^n) \notag
\\& \times \sum_{  \underline s_i \in \mathcal S} P_{ { \underline S}_i |  \underline X_i^L, Y_i^L} ({\underline s}_i |  \underline x_i^L, y_i^L) d( \underline s_i, \hat{ \underline s}_i) \label{eq:condition-on-each-i}
\\& \geq \sum_{\underline x^n, y^n} P_{ \underline X^n, Y^n}( \underline x^n, y^n)  \notag
\\& \times \min_{\hat{ \underline s}_i \in \mathcal S}  \sum_{  \underline s_i \in \mathcal S} P_{ { \underline S}_i | \underline X_i^L, Y_i^L} ({\underline s}_i | \underline x_i^L, y_i^L) d( \underline s_i, \hat{ \underline s}_i) \notag
\\& = \mathbb{E}[d( \underline S_i, \hat{ \underline s}^{\star}( \underline X_i^L, Y_i^L))], 
\end{align}
\end{subequations}
where  \eqref{eq:condition-on-each-i} holds by the Markov chain
\begin{align*}
( \underline X_1^{L}, \cdots,  \underline X_{i-1}^L, Y_1^{L}, \cdots, Y_{i-1}^L, \hat{ \underline S}_i) - (\underline X_i^L, Y_i^L) - \underline S_i.
\end{align*}
Summing over all $i=1, \cdots, \ell$, we have 
\begin{align*}
\Delta^{(\ell)} &= \frac{1}{\ell} \sum_{i=1}^{\ell} \mathbb{E}[ d(  \underline S_i, \hat{ \underline S}_i ) ]
\geq \frac{1}{\ell} \sum_{i=1}^{\ell}\mathbb{E}[d(  \underline S_i, \hat{ \underline s}^{\star}( \underline X_i^L, Y_i^L))], \end{align*}
 which leads to the desired conclusion. 
\end{IEEEproof}

Lemma \ref{lemma-s-star} allows us to define the conditional estimation cost
\begin{align*}
c( \underline x_i^L) \delequal \mathbb{E}[ d(  \underline S_i, \hat{ \underline s}^{\star}( \underline X_i^L, Y_i^L)) | \underline X_i^L =  \underline x_i^L],
\end{align*}
such that, for any encoding function 
\begin{align}
\Delta^{(\ell)}= \frac{1}{\ell} \sum_{i=1}^{\ell} \mathbb{E}[ c( \underline x_i^L)]. \label{eq:distortion-block}
\end{align}

\begin{definition}\normalfont
\label{def:C-Dmax}
Define the minimum distortion $D( B_\text{peak} )$ under the peak input cost constraint $B_\text{peak}$ as
   \begin{align}
\min_{ P_{ { \underline X}^L}  }  
   \frac{1}{\ell} \sum_{i=1}^{\ell}   \sum_{ { \underline x}^L} P_{ {\underline X}^L}( {\underline  x}^L) c( \underline x^L), \label{eq:D-max}
 \end{align}
 where  $P_{ {\underline X}^L}( {\underline x}^L) $ satisfies the peak cost constraint, i.e.,  $b({\underline X}_{i,j}) \leq B_\text{peak}, \forall i\in[\ell], j\in[L]$ where $b(\cdot): \mathcal X \to \mathbb R_+$ is an input cost function.\end{definition}

Since the channel state $\underline S$ is i.i.d. over each block, without loss of generality, we consider only the first block and ignore the block index $i$. The same derivation/strategy can be applied to other blocks identically. We consider $b(\cdot)$ to be the Hamming weight function (number of ones). 
 This is physically motivated by the fact that assuming constant transmission power per direction, the total transmission power is proportional to the number of directions in which $\underline X_{i,j}$ sends a ``1’’. 
The estimation distortion function $d(\underline s, \hat{ \underline s})$ is characterized by Hamming distance, that is,
\begin{align} \label{eq:d}
d(\underline s, \hat{ \underline s}) = \begin{cases}
0, & \underline s = \hat{ \underline s}
\\
2, &   \underline s, \hat{\underline s} \in \mathcal S \text{ and } \underline s \neq \hat{ \underline s} 
\end{cases},
\end{align} 
 since $\underline s$ and $\hat{ \underline s}$ are both one-hot vectors. 


\section{Main Results}
\label{sec:main}
In this section, we first derive the minimum distortion under a peak cost constraint $B_\text{peak}$  with an unconstrained communication of the BBP channel model, i.e., $D(B_\text{peak})$ in \eqref{eq:D-max}. Then, we provide a sensing strategy that achieves the minimum distortion. Notice that this strategy can simultaneously achieve the capacity of this BBP channel by our previous result \cite{Li2022OnTC}.

\subsection{Minimum Distortion}
\label{sub:minimumdistortion}
 Let ${\mathcal B}_{y^j}(\underline x^j)$ denote the set of beam indices containing the transmission direction at channel use $j$ when channel input $\underline X^j = \underline x^j$ and feedback $Y^j = y^j$  for all possible transmission strategies. Then, we can simplify the distortion in \eqref{eq:distortion-block} as follows. We initialize ${\mathcal B}_{y^0}( \underline x^0) = [M]$. The state estimation decision is made based on ${\mathcal B}_{y^L}( \underline x^L)$.  For this BBP channel with iBM and noiseless feedback, we have 
 \begin{align*}
 P_{ \hat{\underline S} | \underline X^L Y^L} ({\underline s} | \underline x^L, y^L) &= P_{ {\underline S} | \underline X^L Y^L} ({\underline s} | \underline x^L, y^L)
 \\& = \frac{ P_{ {\underline S}, \underline X^L Y^L} ({\underline s}, \underline x^L, y^L)}{  P_{  \underline X^L Y^L} (\underline x^L, y^L)}.
 \end{align*}
The joint distribution for $L$ channel uses is
\begin{subequations}
     \begin{align}
     &P_{ {\underline X}^{L}, Y^{L}}({\underline x}^{L}, y^{L} ),  \notag
     \\&= \sum_{\underline s} P_{ {\underline X}^{L}, Y^{L}, {\underline S}}({\underline x}^{L}, y^{L} , {\underline s}) \notag
     \\& = \sum_{\underline s}  \prod_{j=1}^{L} P_{\underline S}( {\underline s}) \mathbbm{1}_{ \{{\underline s}^T {\underline x}_{j} =y_j \}} P_{ {\underline X}_{j} | {\underline X}^{j-1}, Y^{j-1} }( {\underline x}_{j} | {\underline x}^{j-1}, y^{j-1}) \notag
     \\& = \frac{ | \mathcal B_{y^{L}}( {\underline x}^{L} ) | }{M}  \prod_{j=1}^{L} P_{ {\underline X}_{j} | {\underline X}^{j-1}, Y^{j-1} }( {\underline x}_{j} | {\underline x}^{j-1}, y^{j-1}),   \label{eq:joint-xys}
     \\& = \frac{ | \mathcal B_{y^{L}}( {\underline x}^{L} ) | }{M}  P_{ {\underline X}^{L} ||  Y^{L-1} }( {\underline x}^{L} ||  y^{L-1}), \label{eq:PXLYL}
      \end{align} 
      \end{subequations}
    where \eqref{eq:joint-xys} holds since $\prod_{j=1}^{L} \mathbbm{1}_{ \{{\underline s}^T {\underline x}_{j} =y_j \}} =1$ only for the beam indices belonging to $ \mathcal B_{y^{L} }( {\underline x}^{L})$ and \eqref{eq:PXLYL} holds since we define 
  \begin{align}
  P_{ {\underline X}^{L} | | Y^{L-1} }( {\underline x}^{L} || y^{L-1}) \delequal \prod_{j=1}^L P_{\underline X_j| \underline X^{j-1}, Y^{j-1}} (\underline x_j| \underline x^{j-1}, y^{j-1} ). \label{def:PX^L||Y^L-1}
   \end{align} 
 Therefore, 
 \begin{align}
 P_{ \hat{\underline S} | \underline X^L Y^L} ({\underline s} | \underline x^L, y^L) &= P_{ {\underline S} | \underline X^L Y^L} ({\underline s} | \underline x^L, y^L) \notag
 \\& = \begin{cases}
  \frac{1}{ |{\mathcal B}_{y^L}( \underline x^L)|}, & \forall \underline s\in {\mathcal B}_{y^L}( \underline x^L)
  \\0, & \text{otherwise}
  \end{cases},
  \label{eq:hat-s-prob}
\end{align} 
where $ |{\mathcal B}_{y^L}( \underline x^L)| \geq 1$, i.e., it is uniform over the restricted set $B_y^L( \underline x^L)$ and zero elsewhere.
The distortion can be simplified as
\begin{subequations}
 \begin{align}
&\mathbb{E}[d( \underline S, \hat{ \underline S})] \notag
\\&= \mathbb{E}_{ \underline X^L, Y^L} \left[ \mathbb{E}[d( \underline S, \hat{\underline S}) | \underline X^L, Y^L] \right] \notag
\\ & = \sum_{\underline x^L, y^L} P_{\underline X^L Y^L}(\underline x^L, y^L) \sum_{\hat{\underline s} \in \mathcal S} P_{ \hat{\underline S} | \underline X^L Y^L} (\hat{\underline s} | \underline x^L, y^L) \notag
\\& \times \sum_{ \underline s \in \mathcal S} P_{ {\underline S} | \underline X^L Y^L} ({\underline s} | \underline x^L, y^L) d( \underline s, \hat{\underline s}) \notag
\\& = \sum_{\underline x^L, y^L} P_{\underline X^L Y^L}(\underline x^L, y^L)  \frac{2 [ |{\mathcal B}_{y^L}( \underline x^L)| -1]^+}{|{\mathcal B}_{y^L}( \underline x^L)|}  \label{eq:condition-on-each-i-det}
 \\& = \sum_{ {\underline x}^{L}}  \sum_{y^{L}}  \prod_{j=1}^{L} P_{ {\underline X}_{j} | {\underline X}^{j-1}, Y^{j-1} }( {\underline x}_{j} | {\underline x}^{j-1}, y^{j-1})  \frac{ | \mathcal B_{y^{L}}( {\underline x}^{L} ) | }{M} \notag
 \\& \qquad \times \frac{ 2[| \mathcal B_{y^{L}}( {\underline x}^{L} ) |  -1]^+ }{| \mathcal B_{y^{L}}( {\underline x}^{L} ) |}
\notag 
 \\& =  \sum_{ {\underline x}^{L}}  \sum_{y^{L}}  P_{ {\underline X}^{L} ||  Y^{L-1} }( {\underline x}^{L} || y^{L-1})  \frac{ 2 [ | \mathcal B_{y^{L}}( {\underline x}^{L} ) | -1 ]^+}{M}  \label{eq:simplified-D}
\end{align} 
\end{subequations}
where \eqref{eq:condition-on-each-i-det} follows from \eqref{eq:d} and \eqref{eq:hat-s-prob}, and \eqref{eq:simplified-D} follows from \eqref{def:PX^L||Y^L-1}. 

Sending back a $Y_k =1$ indicates that the transmission direction is detected within the small set of ones in $\underline X_{k}$. Recall that $\beta_k^L$ denotes the set containing all possible $L$-length binary sequences with the first non-zero element appearing at index $k$. 
 Let 
 \begin{subequations}
      \label{eq:peak-prob-notations}
 \begin{align}
c_k \delequal M\sum_{y^L \in \beta_k^L} P_{Y^L}(y^L), \label{def:ck}
\end{align}
 which  
 is independent of the transmission strategy.
   Then,  \begin{align}
     P_{Y^L}(0^L) & = 1 - \sum_{k=1}^{L }  \sum_{y^L \in \beta_k^L} P_{Y^L}(y^L) 
 =1 - \sum_{k=1}^{L } \frac{c_k}{M}. \label{def:ck-0}
     \end{align}
   Further, by \eqref{def:ck}, we have 
   \begin{align}
   c_k \leq M P_{Y_k}(y_k = 1) \leq M \frac{B_\text{peak}}{M} = B_\text{peak}, \label{eq:peak-ck}
      \end{align}
      where \eqref{eq:peak-ck} holds by the peak input cost constraint. 
     \end{subequations}
 Following this notation, we next provide the minimum distortion under the peak cost constraint. 

\begin{theorem} \label{thm:min-dis}
The minimum distortion $D(B_\text{peak})$  defined in \eqref{eq:D-max} of the BBP model with iBM under peak cost constraint $B_\text{peak}$ is 
     \begin{align}
  & D(B_\text{peak}) = \sum_{j=1}^L  \frac{ 2 [ c_j -2^{L-j} ]^+ }{ M } +  \frac{2[ M - \sum_{j=1}^{L} c_{j} -1]^+}{M },\label{eq:peak-each-d}
  \end{align} 
where  $c_1 = \min( \frac{M}{2}, B_\text{peak})$, and
   \begin{align}
    c_j = \min( \frac{M- \sum_{k=1}^{j-1} c_k }{2}, B_\text{peak} ), 1 <j \leq L.\label{eq:b0s-peak}
  \end{align} 
\end{theorem}
\begin{IEEEproof}
We first prove that the minimum distortion is presented in \eqref{eq:peak-each-d}. 
 By \eqref{def:ck}, we have
   \begin{align}
 c_{j} & = M \sum_{\underline x^L} \sum_{y^L \in \beta_j^L}P_{\underline X^L, Y^L} (\underline x^L, y^L ) \notag  
 \\& =  M \sum_{\underline x^L}  \sum_{y^L \in \beta_j^L} \frac{ | \mathcal B_{y^L}(\underline x^L)|}{M} P_{ {\underline X}^{L} | | Y^{L-1} }( {\underline x}^{L} || y^{L-1}) \label{eq:cj-B}
 \end{align}
 where \eqref{eq:cj-B} follows from \eqref{eq:PXLYL}. Similarly, by \eqref{def:ck-0}, we have
  \begin{align*}
 M - \sum_{j=1}^L c_{j} 
 & = M \sum_{\underline x^L} P_{\underline X^L, Y^L} (\underline x^L, 0^L ) \notag 
 \\& = M \sum_{\underline x^L}  \frac{ | \mathcal B_{0^L}(\underline x^L)|}{M}  P_{\underline X^L||  Y^{L-1}} (\underline x^L|| \underline 0^{L-1} ).
 \end{align*}
Continuing with \eqref{eq:simplified-D},
the distortion at the end of each block is at least
 \begin{subequations}
   \begin{align}
   &D  \notag
 \\& =  \sum_{j=1}^L \sum_{y^L \in \beta_j^L}  \sum_{ {\underline x}^{L}}  P_{ {\underline X}^{L} | | Y^{L-1} }( {\underline x}^{L} || y^{L-1})  \frac{ 2 [ | \mathcal B_{y^{L}}( {\underline x}^{L} ) | - 1 ]^+}{M}  \notag
 \\& +  \sum_{ {\underline x}^{L}}  P_{ {\underline X}^{L} | | Y^{L-1} }( {\underline x}^{L} || 0^{L-1})  \frac{ 2 [ | \mathcal B_{0^{L}}( {\underline x}^{L} ) | - 1 ]^+}{M} \notag
 \\& \! \geq  \!  \sum_{j=1}^L \! \frac{ 2 [ \sum_{y^L \in \beta_j^L}  (\sum_{ {\underline x}^{L}} \!  P_{ {\underline X}^{L} | | Y^{L-1} }( {\underline x}^{L} || y^{L-1}) | \mathcal B_{y^{L}}( {\underline x}^{L} ) | -1) ]^+}{M} \label{eq:convex}
 \\& + \frac{ 2 [  \sum_{ {\underline x}^{L}}   P_{ {\underline X}^{L} | | Y^{L-1} }( {\underline x}^{L} || 0^{L-1}) | \mathcal B_{0^{L}}( {\underline x}^{L} ) | -1 ]^+}{M} \notag
 \\& =  \sum_{j=1}^L  \frac{ 2 [ c_j  -2^{L-j} ]^+}{M} +  \frac{ 2 [ M - \sum_{j=1}^L c_j  -1  ]^+}{M} \label{eq:peak-D-result}
    \end{align}
    \end{subequations} 
    where \eqref{eq:convex} holds since $[x-1]^+$ is a convex function, and \eqref{eq:peak-D-result} holds  by \eqref{eq:cj-B} and since there are $2^{L-j}$ possible $y^L$ belongs to $\beta_j^L$. Hence, the minimum distortion can be represented by \eqref{eq:peak-each-d}. 

Next, we show that the minimum \eqref{eq:peak-each-d} can be obtained by choosing $c_j, i\in[L]$ iteratively as given in \eqref{eq:b0s-peak}.   Ideally, the minimum of \eqref{eq:peak-D-result}  is achieved when 
%
$M-\sum_{j=1}^L c_j -1 \geq 0$
and $ c_j - 2^{L-i} \geq 0, \forall j \in [L]$. 
Hence, we have 
$   M - 1 \geq \sum_{j=1}^L 2^{L-j},$
       which gives $L\leq \log M$.  To have $c_j \geq 2^{L-j} $ and  $M-\sum_{j=1}^L c_j -1 \geq 0$ hold simultaneously, we can choose $c_1 = \frac{M}{2}$ and $c_j = \frac{M-\sum_{k=1}^{j-1} c_k}{2}$. Meanwhile, by \eqref{eq:peak-ck}, $c_j \leq B_\text{peak}$ for all $j \in[L]$. Therefore, we can choose \eqref{eq:b0s-peak} to achieve the minimum of \eqref{eq:peak-each-d}. Similarly, one can verify that the minimum distortion $ D(B_\text{peak}) $ is achieved by choosing \eqref{eq:b0s-peak} when $L > \log M$. 
\end{IEEEproof}

 \begin{remark}
 \normalfont
The optimal choice of $c_j$ to minimize \eqref{eq:peak-each-d} is not unique and depends on the values of $L, M$ and $B_\text{peak}$. For example, when $L=1, B_\text{peak} >1,$ and $M>1$, any $ 1 \leq c_i \leq M-1$ achieves the minimum distortion. Herein, we choose $c_j$ as in \eqref{eq:b0s-peak} since it also achieves the channel capacity as proved in \cite{Li2022OnTC}.
\end{remark}

\subsection{Estimation Strategy}
In order to minimize distortion, it is critical to reducing the size of the set $\mathcal B_{y^L}(\underline x^L)$.   
 Let $\mathcal B_i^e$ denote the set of beam indices to be explored at channel use $i$ and $\mathcal B_i^{e,c}$ denote the complementary of $\mathcal B_i^e$ (i.e., $\mathcal B_i^e \cup \mathcal B_i^{e,c} = [M]$). Initially, ${\mathcal B}_{y^0} = [M]$ and ${\mathcal B}_{0}^e = \emptyset$. 
  Based on the strictly causal noiseless feedback,  $|\mathcal B_{y^i}(\underline x^i)|$ can be updated as
\begin{align}
&| \mathcal B_{y^{i+1}}(\underline x^{i+1} ) | \notag
\\& = y_{i+1}  | \mathcal B_{y^{i}}(\underline x^{i} ) \cap \mathcal B_i^e| + (1 - y_{i+1}) | \mathcal B_{y^{i}}(\underline x^{i} ) \cap \mathcal B_i^{e,c}| \label{eq:transmission-set-update}
\\& \geq y_{i+1}  | \mathcal B_i^e| + (1 - y_{i+1}) (| \mathcal B_{y^{i}}(\underline x^{i} ) | -| \mathcal B_i^{e}|) \label{eq:bound-transmission-set}
\end{align}
where \eqref{eq:transmission-set-update} indicates that the size of possible transmission directions is decreasing (i.e., $| \mathcal B_{y^{i+1}}(\underline x^{i+1} ) | \leq | \mathcal B_{y^{i}}(\underline x^{i} ) |$) and 
equality in \eqref{eq:bound-transmission-set} holds when $  \mathcal B_i^e \subseteq B_{y^{i}}(\underline x^{i} )$, that is, the transmitter selects beam indices from the set $B_{y^{i}}(\underline x^{i} )$ recursively. 

Following the ideas illustrated above, we next show that the minimum distortion in Theorem \ref{thm:min-dis} can be obtained by applying the transmission strategy in \cite[Algorithm 1]{Li2022OnTC}. Specifically, we initialize a sequence of $\{ c_1, \cdots, c_L\}$ iteratively solved by \eqref{eq:b0s-peak}. At the beginning of channel use $i$, we update ${\mathcal B}_{y^i}$ and choose some number of beam indices randomly and uniformly from ${\mathcal B}_{y^{i}}$ based on the casual feedback $Y_{i-1}$. 
Additionally, we use $k, k\in[L]$ to record the number of channel uses until the transmitter selected the ``right" directions (i.e., $Y_{k}=1$). Before that, the transmitter randomly and uniformly chooses $c_i, i\leq k$ beam indices from ${\mathcal B}_{y^{i-1}}$. After that,  the transmitter randomly and uniformly chooses $\frac{c_k}{2^{i-k}}, i>k$ beam indices from ${\mathcal B}_{y^{i-1}}$. These selected beam indices are stored in set $\mathcal B_i^e$. 


Recall that $\beta_{k}^L$ denotes the binary sequence with length $L$ and the first non-zero element appearing at the $k$-index. Based on this transmission strategy,  the probabilities of output sequences $y^L$ under the condition of different channel states are the same, i.e., $P_{Y^L | \underline S}(y^L | \underline s) = P_{Y^L | \underline S}(y^L | \underline s^{\prime}),  \underline s \neq \underline s^{\prime}$, and one can easily check that 
 \begin{subequations}
 \label{subeq:assumption-di}
  \begin{align}
& \sum_{\underline x^L} \sum_{y^L \in \beta_k^{L} }P_{ {\underline X}^{L}, Y^{L}}({\underline x}^{L},y^L) = \frac{ c_k}{M}, 
\  | \mathcal B_{\beta_k^{L}}( {\underline x}^L ) | = \frac{c_k }{2^{L-k}}, \label{eq:betaj-i}
 \\
&  \sum_{\underline x^L} P_{ {\underline X}^{L}, Y^{L}}({\underline x}^{L}, 0^{L} ) = 1- \frac{\sum_{k=1}^L c_k }{M},\ | \mathcal B_{0^{L}}( {\underline x}^{L} ) | =  M - \sum_{k=1}^L c_k, \label{eq:remaining-set}
 \end{align} 
 \end{subequations}
 where $ | \mathcal B_{0^{L}}( {\underline x}^{L} ) |  \geq 0$ by \eqref{eq:b0s-peak} and $\mathcal B_{\beta_k^{L}}( {\underline x}^L )$ denotes the set containing possible transmission directions for any channel input sequence ${\underline x}^L$ leading to $y^L \in \beta_k^{L}$. From  \eqref{eq:simplified-D}, the distortion is at least
      \begin{subequations}
      \label{subeqs:peak-d}
   \begin{align}
   &D(B_\text{peak}) 
 = \sum_{ {\underline x}^{L}}  \left( \sum_{k=1}^{L} \sum_{y^{L} \in \beta_k^{L}}
    \right. \notag
 \\&  \left.  P_{ {\underline X}^{L} ||  Y^{L-1} }( {\underline x}^{L} ||  y^{L-1} ) \frac{ 2 [ | \mathcal B_{ \beta_k^L }( {\underline x}^{L} ) | -1 ]^+}{M} \right.  \label{eq:replace-y^i-beta}
 \\& \left.  +   P_{ {\underline X}^{L} ||  Y^{L-1} }( {\underline x}^{L} ||  0^{L-1})   \frac{ 2 [ | \mathcal B_{0^{L}  }( {\underline x}^{L} ) | -1 ]^+}{M}  \right)  \label{eq:replace-y^i-0^i}
 \\& =   \sum_{k=1}^{L}  2^{L-k}  \frac{2 [\frac{c_k }{2^{L-k}}  -1 ]^+ }{M }  +  \frac{2 [ M - \sum_{k=1}^L c_k  -1 ]^+ }{M }
  \label{eq:distortion-lower-bound-0-i+1}
 %
         \\& = \sum_{k=1}^{L} \frac{ 2 [ c_k - 2^{L-k}]^+}{M} + \frac{ 2 [ M - \sum_{k=1}^{L} c_{k}   - 1]^+}{M}. \label{eq:disortion-final}
             \end{align}  
 \end{subequations} 
We partition the sequence of $y^{L}$ into $ y^L \in \beta_k^{L}$ and $y^L = 0^{L}$ in \eqref{eq:replace-y^i-beta} and \eqref{eq:replace-y^i-0^i}.
\eqref{eq:distortion-lower-bound-0-i+1} follows from \eqref{eq:betaj-i} and there are $2^{i-k}$ possible $y^i$ sequences in $\beta_k^i$ sharing the same probability $P_{ {\underline X}^{i}, Y^{i}}({\underline x}^{i}, y^i \in \beta_{k}^{i} )$ for $k\leq i$ according to  Algorithm 1.
 Finally, we obtain the lower bound \eqref{eq:disortion-final}.  Therefore, we showed that  Algorithm 1 in \cite{Li2022OnTC} achieves the distortion in \eqref{eq:peak-each-d}.


\section{Conclusion}

\label{sec:conclude}
In this work, we studied a binary beam-pointing channel with in-block memory and feedback that captures the main feature of the beam alignment problem in mmWave communications and yet is sufficiently simple to be tractable from an information-theoretic viewpoint. 
We derived the minimum distortion of this simplified channel model under the peak cost constraint. We showed that the capacity-achieving transmission strategy in \cite{Li2022OnTC} attains the minimum distortion simultaneously. In conclusion, we have characterized the full capacity-distortion region of this binary beam-pointing channel under the peak cost constraint. This surprising result reveals the fact that channel estimation and signal communication can be jointly optimal, which enables the efficient utilization of the available resources in time, frequency, available antennas, and transmission power.    

 

\bibliographystyle{IEEEtran}
\bibliography{refs-isac}

\begin{thebibliography}{10}
\providecommand{\url}[1]{#1}
\csname url@samestyle\endcsname
\providecommand{\newblock}{\relax}
\providecommand{\bibinfo}[2]{#2}
\providecommand{\BIBentrySTDinterwordspacing}{\spaceskip=0pt\relax}
\providecommand{\BIBentryALTinterwordstretchfactor}{4}
\providecommand{\BIBentryALTinterwordspacing}{\spaceskip=\fontdimen2\font plus
\BIBentryALTinterwordstretchfactor\fontdimen3\font minus
  \fontdimen4\font\relax}
\providecommand{\BIBforeignlanguage}[2]{{%
\expandafter\ifx\csname l@#1\endcsname\relax
\typeout{** WARNING: IEEEtran.bst: No hyphenation pattern has been}%
\typeout{** loaded for the language `#1'. Using the pattern for}%
\typeout{** the default language instead.}%
\else
\language=\csname l@#1\endcsname
\fi
#2}}
\providecommand{\BIBdecl}{\relax}
\BIBdecl

\bibitem{Kramer-memory}
G.~Kramer, ``{Information Networks With In-Block Memory},'' \emph{IEEE
  Transactions on Information Theory}, vol.~60, no.~4, pp. 2105--2120, April
  2014.

\bibitem{Li2022OnTC}
S.~Li and G.~Caire, ``{On the Capacity of “Beam-Pointing” Channels with
  Block Memory and Feedback: The Binary Case},'' \emph{2022 56th Asilomar
  Conference on Signals, Systems, and Computers}, pp. 1262--1268, 2022.

\bibitem{OverviewmmWave}
R.~W. Heath, N.~González-Prelcic, S.~Rangan, W.~Roh, and A.~M. Sayeed, ``{An
  Overview of Signal Processing Techniques for Millimeter Wave MIMO Systems},''
  \emph{IEEE Journal of Selected Topics in Signal Processing}, vol.~10, no.~3,
  pp. 436--453, 2016.

\bibitem{5G-fixed-wireless-access}
K.~Aldubaikhy, W.~Wu, N.~Zhang, N.~Cheng, and X.~Shen, ``{mmWave IEEE 802.11ay
  for 5G Fixed Wireless Access},'' \emph{IEEE Wireless Communications},
  vol.~27, no.~2, pp. 88--95, 2020.

\bibitem{SensorIngelligence}
A.~K.~R. Chavva, S.~K., C.~Lim, Y.~Lee, J.~Kim, and Y.~Rashid, ``Sensor
  intelligence based beam tracking for 5g mmwave systems: A practical
  approach,'' in \emph{2019 IEEE Global Communications Conference (GLOBECOM)},
  2019, pp. 1--6.

\bibitem{Location-aware}
I.~Orikumhi, J.~Kang, C.~Park, J.~Yang, and S.~Kim, ``{Location-Aware
  Coordinated Beam Alignment in mmWave Communication},'' in \emph{2018 56th
  Annual Allerton Conference on Communication, Control, and Computing
  (Allerton)}, 2018, pp. 386--390.

\bibitem{mmWave5G}
W.~Roh, J.-Y. Seol, J.~Park, B.~Lee, J.~Lee, Y.~Kim, J.~Cho, K.~Cheun, and
  F.~Aryanfar, ``{Millimeter-wave beamforming as an enabling technology for 5G
  cellular communications: theoretical feasibility and prototype results},''
  \emph{IEEE Communications Magazine}, vol.~52, no.~2, pp. 106--113, 2014.

\bibitem{BeamDetection}
M.~S. Ullah and A.~Tewfik, ``{Beam Detection Analysis for 5G mmWave Initial
  Acquisition},'' in \emph{2018 28th International Telecommunication Networks
  and Applications Conference (ITNAC)}, 2018, pp. 1--8.

\bibitem{Desai_2014}
V.~Desai, L.~Krzymien, P.~Sartori, W.~Xiao, A.~C.~K. Soong, and A.~Alkhateeb,
  ``{Initial beamforming for mmWave communications},'' \emph{null}, 2014.

\bibitem{Javidi-AL}
S.-E. Chiu, N.~Ronquillo, and T.~Javidi, ``{Active Learning and CSI Acquisition
  for mmWave Initial Alignment},'' \emph{IEEE Journal on Selected Areas in
  Communications}, vol.~37, no.~11, pp. 2474--2489, 2019.

\bibitem{Hashemi_2018}
M.~Hashemi, A.~Sabharwal, C.~E. Koksal, and N.~B. Shroff, ``{Efficient Beam
  Alignment in Millimeter Wave Systems Using Contextual Bandits},''
  \emph{null}, 2018.

\bibitem{Michelusi_2018}
N.~Michelusi and M.~Hussain, ``{Optimal Beam-Sweeping and Communication in
  Mobile Millimeter-Wave Networks},'' \emph{null}, 2018.

\bibitem{Scalable-song}
X.~Song, S.~Haghighatshoar, and G.~Caire, ``{A Scalable and Statistically
  Robust Beam Alignment Technique for Millimeter-Wave Systems},'' \emph{IEEE
  Transactions on Wireless Communications}, vol.~17, no.~7, pp. 4792--4805,
  2018.

\bibitem{mmPosterior}
F.~Pedraza and G.~Caire, ``{Adaptive Two-Sided Beam Alignment in mmWave via
  Posterior Matching},'' in \emph{2020 IEEE Information Theory Workshop (ITW)},
  2021, pp. 1--5.

\bibitem{Hussain_2019}
M.~Hussain and N.~Michelusi, ``{Energy-Efficient Interactive Beam Alignment for
  Millimeter-Wave Networks},'' \emph{IEEE Transactions on Wireless
  Communications}, 2019.

\bibitem{Lapidoth1998}
A.~Lapidoth and P.~Narayan, ``{Reliable communication under channel
  uncertainty},'' \emph{IEEE Transactions on Information Theory}, vol.~44,
  no.~6, pp. 2148--2177, 1998.

\bibitem{ISAC-6G}
F.~Liu, Y.~Cui, C.~Masouros, J.~Xu, T.~X. Han, Y.~C. Eldar, and S.~Buzzi,
  ``{Integrated Sensing and Communications: Towards Dual-functional Wireless
  Networks for 6G and Beyond},'' \emph{IEEE Journal on Selected Areas in
  Communications}, pp. 1--1, 2022.

\bibitem{ISAC-Tradeoff}
\BIBentryALTinterwordspacing
M.~Kobayashi, G.~Caire, and G.~Kramer, ``{Joint State Sensing and
  Communication: Optimal Tradeoff for a Memoryless Case},'' 2018. [Online].
  Available: \url{https://arxiv.org/abs/1805.05713}
\BIBentrySTDinterwordspacing

\bibitem{ISAC-MAC}
M.~Kobayashi, H.~Hamad, G.~Kramer, and G.~Caire, ``{Joint State Sensing and
  Communication over Memoryless Multiple Access Channels},'' in \emph{2019 IEEE
  International Symposium on Information Theory (ISIT)}, 2019, pp. 270--274.

\bibitem{ISAC-IT}
M.~Ahmadipour, M.~Kobayashi, M.~Wigger, and G.~Caire, ``An
  information-theoretic approach to joint sensing and communication,''
  \emph{IEEE Transactions on Information Theory}, pp. 1--1, 2022.

\end{thebibliography}
\end{document}